\documentclass[prl,twocolumn,showpacs,preprintnumbers,superscriptaddress,amsmath,amssymb]{revtex4}
\usepackage{graphicx}
\usepackage{dcolumn}
\usepackage{bm}

\begin{document}

\title{Dynamic interference of photoelectrons produced by high-frequency laser pulses}

\author{\firstname{Philipp~V.} \surname{Demekhin}}
\altaffiliation{philipp.demekhin@pci.uni-heidelberg.de}
\affiliation{Theoretische Chemie, Physikalisch-Chemisches
Institut, Universit\"{a}t  Heidelberg, Im Neuenheimer Feld 229,
D-69120 Heidelberg, Germany}
\affiliation{Rostov State Transport University, Narodnogo Opolcheniya square 2,  Rostov-on-Don, 344038, Russia}

\author{\firstname{Lorenz~S.} \surname{Cederbaum}}
\altaffiliation{lorenz.cederbaum@pci.uni-heidelberg.de}
\affiliation{Theoretische Chemie, Physikalisch-Chemisches Institut, Universit\"{a}t Heidelberg,
Im Neuenheimer Feld 229, D-69120 Heidelberg, Germany}
\date{\today}

\begin{abstract}
The ionization of an atom by a high-frequency intense laser pulse, where the energy of a single-photon is sufficient to ionize the system, is investigated from first principles. It is shown that as a consequence of an AC Stark effect in the continuum, the energy of the photoelectron follows the envelope of the laser pulse. This is demonstrated to result in strong dynamic interference of the photoelectrons of the same kinetic energy emitted at different times. Numerically exact computations on the hydrogen atom demonstrate that the dynamic interference spectacularly modifies the photoionization process and is prominently manifested in the photoelectron spectrum by the appearance of a distinct multi-peak pattern. The general theory is  well approximated by explicit analytical expressions which allow for a transparent understanding of the discovered phenomena and for making predictions on the dependence of the measured spectrum on the  pulse.
\end{abstract}

\pacs{33.20.Xx,  41.60.Cr, 82.50.Kx}
\maketitle

Dipole transitions between bound electronic states of matter mediated by intense laser pulses are affected by a fundamental energy shift induced by the oscillating external field of the laser known as the dynamic or AC Stark effect \cite{B1,B2,B3,B4,A1,A2,A3,A4}. The AC Stark effect induced by optical laser fields is used in a myriad of contemporary experiments to hold and align molecules \cite{align}, to shape potential energy surfaces  \cite{shape}, to make rapid transient birefringence \cite{birefringe1,birefringe2}, and to control quantum systems \cite{control1,control2}. What happens if the transition is to the continuum and not to a bound electronic state? Here, we investigate the photoionization of an atom by a high-frequency strong laser pulse, where the energy of a single-photon is sufficient to ionize the system. We demonstrate  that the interplay between the photoionization and the AC Stark shift results in a hitherto unrecognized dynamic interference of photoelectrons of the same kinetic energy emitted at different times. The effect is universal and  prominently manifested in the photoelectron spectrum.

In bound-bound electronic transitions accessible to optical lasers, the AC Stark effect detunes the laser frequency away from the field-free resonant frequency, an effect which has to be compensated in the presence of strong fields in order to have a transition. The situation is different in the case of ionization. The new generation of light sources, like the free electron laser (FEL) at FLASH \cite{FLASH}, and recent progress in high-order harmonic generation techniques \cite{highharm1,highharm2} allow one to produce ultrashort laser pulses with single-photon energies well above the ionization threshold of any matter. This immediately raises the fundamental question of how the well-studied single-photon ionization process in weak fields will be modified in intense laser fields of high frequencies. We shall demonstrate here that single-photon ionization by high-frequency strong pulses is accompanied by dynamic interference caused by the AC Stark effect in the continuum.

In order to illustrate the new fundamental effect, we have chosen as an explicit example the photoionization of the hydrogen atom which is not affected by many-electron correlations and, thus, is amenable to exact computations. The AC Stark effect arises from the indirect coupling of nonresonant (nonessential) states that do not participate directly in the excitation process  \cite{Sussman11}. To arrive at an unequivocal description of the process, we have computed exactly the quantum motion of the whole discrete and continuum electron spectrum of hydrogen exposed to a strong
pulse. For the sake of transparency of presentation we show below explicitly only the essential states (i.e., $1s\to \varepsilon p$ ionization) and employ the simplifying rotating wave approximation (RWA), which turns out to describe well the underlying physics. In the full calculations we have used also the nonessential states and discarded the RWA (see below and Appendix for details).

An atom initially in its ground electronic state $\vert I\rangle$ of energy of $E_I=0$ (chosen as the origin of the energy scale) is ionized  into the final electron continuum state $\vert F \varepsilon\rangle$ of energy $IP + \varepsilon$ (where $IP=E_F-E_I$ is  the ionization potential and $\varepsilon$ is the kinetic energy of the photoelectron) by a linearly polarized laser pulse of carrier frequency $\omega$ and pulse-shape function $g(t)$.  Following \cite{Pahl99ZPD,Chiang10,Demekhin11SFatom,MolRaSfPRL,DemekhinCO,DemekhinICD}, the total wave function of the system as a function of time reads
\begin{equation}
\label{eq:anzatz}
\Psi(t)= a_I(t)\vert I \rangle +\int  d\varepsilon \,a_\varepsilon (t) \vert  F\varepsilon \rangle e^{-i\omega t},
\end{equation}
where $a_I(t)$ and $a_\varepsilon (t)$ are the time-dependent amplitudes of the populations of the initial and final continuum states, the latter being dressed by the field (redefined by multiplying with the phase factor  $e^{i\omega t}$ \cite{Demekhin11SFatom}). Inserting $\Psi(t)$ into the time-dependent Schr\"{o}dinger equation for the total Hamiltonian of the atom plus its interaction with the laser field, we obtain the following set of equations for the amplitudes (atomic units are used throughout)
\begin{subequations}
\label{eq:CDE}
\begin{equation}
\label{eq:CDE_I}
i\dot{a}_I(t)= \int d\varepsilon \begin{array}{c} \left\{\frac{1}{2} d^\dag_\varepsilon \mathcal{E}_0 \right\} \end{array}g(t) \, {a}_\varepsilon(t),
\end{equation}
\begin{equation}
\label{eq:CDE_F}
i\dot{ {a}}_\varepsilon (t)=  \begin{array}{c} \left\{\frac{1}{2} d_\varepsilon \mathcal{E}_0 \right\} \end{array}g(t) \,  a_I(t) + \left(IP+\varepsilon-\omega\right) {a}_\varepsilon (t).
\end{equation}
\end{subequations}
$\mathcal{E}_0$ is the peak amplitude of the field $\mathcal{E}(t)=\mathcal{E}_0 \,g(t) \cos\omega t$, and $d_\varepsilon=\langle F\varepsilon \vert \hat{z}\vert I\rangle$  is the energy-dependent ionization dipole transition matrix element, which can be exactly computed for hydrogen.

\begin{figure}
\includegraphics[scale=0.43]{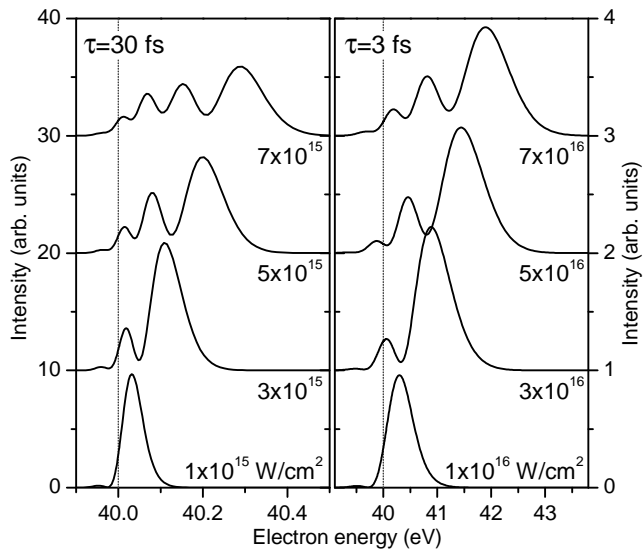}
\caption{Photoelectron spectra of the hydrogen atom exposed to Gaussian-shaped pulses of 30~fs (left panel) and 3~fs (right panel) durations with carrier frequency  $\omega=53.6057$~eV computed exactly for different peak intensities. In weak-field single-photon ionization, the photoelectron line appears as a Gaussian peak centered around the energy of $\varepsilon_0=\omega-IP=40$~eV (indicated by the vertical broken line) and a width provided by the pulse: FWHM of about 52~meV for the 30~fs pulse, and 0.52~eV for the 3~fs pulse. The presently computed strong pulse ionization spectra are, however, much broader (by several times), shifted to higher energies from the central electron energy of $\varepsilon_0=40$~eV, and posses distinct modulations of the electron intensity. It will be shown that these modulations are due to dynamic interference in the continuum.}\label{fig_spectra}
\end{figure}

The calculations were performed for a Gaussian-shaped pulse, $g(t)=e^{-t^2/\tau^2}$,  and carrier frequency $\omega=53.6057$~eV, which is well above the ionization potential $IP=13.6057$~eV. The photoelectron spectra computed for two pulse durations and different peak intensities (defined as $I= {\mathcal{E}^2_0}/{8\pi\alpha}$) are depicted in  Fig.~\ref{fig_spectra}. One can see that the effect of a strong laser pulse on the spectra is enormous. Instead of a single peak at $\varepsilon_0=\omega-IP=40$~eV with the width of the pulse as naively expected from weak field calculations, the spectra are substantially shifted and spread, and posses distinct modulations of the intensity. The effect is so big that it can easily be verified experimentally.

\begin{figure}
\includegraphics[scale=0.4]{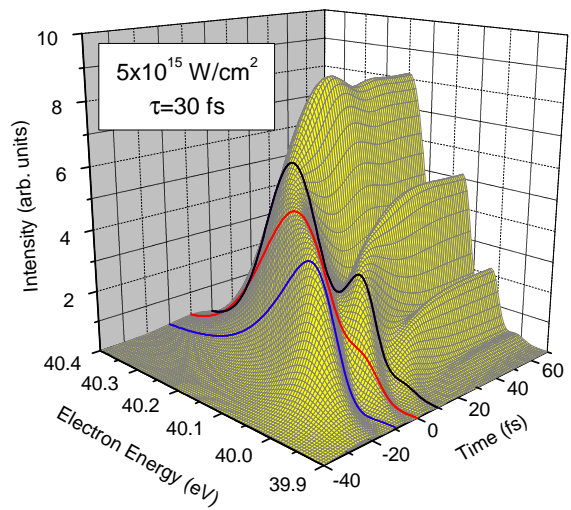}
\caption{(Color online) Time evolution of the photoelectron spectrum computed exactly for the hydrogen atom exposed to a Gaussian-shaped pulse of 30 fs duration, of  carrier frequency  $\omega=53.6057$~eV and peak intensity of $5\times 10^{15}$~W/cm$^2$. At very  early times, the spectrum starts to develop symmetrically around the central electron energy of $\varepsilon_0=40$~eV. At later times, but before the maximum of the pulse arrives, the maximum of the spectrum moves to higher electron energies (see the spectrum computed at --10~fs). When the pulse arrives (the time 0~fs corresponds to the pulse maximum), the spectrum is not any more symmetric, but still does not possess intensity modulations. The first modulations start to show up in the spectrum only after the pulse maximum (see the spectrum computed at +10~fs) has arrived.}\label{fig_time}
\end{figure}

To uncover the underlying physics, we show the time evolution of the photoelectron spectrum in Fig.~\ref{fig_time} computed for the 30~fs pulse and peak intensity of $5\times 10^{15}$~W/cm$^2$. At short times the spectrum develops symmetrically around $\varepsilon_0=40$~eV as expected for weak fields. Until the maximum of the pulse  arrives the spectrum remains nearly symmetric, but its maximum continuously shifts to higher energies. Importantly, strong modulations of the intensity start to develop in the spectrum only after the maximum of the pulse has passed. This will be explained below.

So far, we discussed the spectra computed exactly. Now, we would like to interpret the findings by introducing approximations which allow us to arrive at explicit analytic expressions and proven to be surprisingly accurate. In the local approximation  \cite{Cederbaum81,Domcke91},  Eq.~(\ref{eq:CDE_I}) reads
\begin{equation}
\label{eq:local_I}
i\dot{a}_I(t)= \begin{array}{c}\left( \Delta-\frac{i}{2}\Gamma \right)\end{array}  g^2(t)\,{a}_I(t).
\end{equation}
Here, the real term, $\Delta \,g^2(t)$, is nothing but the AC Stark shift of the energy of the ground state immersed into the dressed continuum. As usual, this shift follows the intensity envelope of the field \cite{Sussman11}, i.e. $g^2(t)$. Derivations of the expression for $\Delta$  can be found, e.g. in \cite{Sussman11}, and in the present case it explicitly reads (without  using the RWA)
\begin{equation}
\label{eq:delta}
\Delta = -\mathcal{P} \int d\varepsilon  \left|  \frac{ d_{\varepsilon }  \mathcal{E}_0}{2}  \right|^2 \left(\frac{1}{IP+\varepsilon -\omega}+\frac{1}{IP+\varepsilon +\omega} \right) ,
\end{equation}
where $\mathcal{P}$ stands for the principal value of the integral. In the RWA approximation the second term on the r.h.s. of Eq.~(\ref{eq:delta}) is neglected. The value of $\Delta$ is proportional to the peak intensity, $\Delta  \sim \mathcal{E}^2_0 \sim I$, and it is essentially nonzero owing to the energy dependence of the dipole matrix element $d_{\varepsilon }$.

The imaginary term  $-\frac{i}{2}\Gamma    g^2(t)$ in Eq.~(\ref{eq:local_I}) describes the losses of the population of the ground state by the ionization into all final electron continuum states $\vert F \varepsilon\rangle$.  $\Gamma$ can be obtained from $\Delta$  in Eq.~(\ref{eq:delta}). The explicit expression for $\Gamma$ was obtained in  \cite{Demekhin11SFatom} in the local and RWA approximations and in the present case it reads
\begin{equation}
\label{eq:gamma}
\Gamma = 2\pi \begin{array}{c}\vert  \frac{1}{2} d_{\varepsilon_0} \mathcal{E}_0 \vert^2 \end{array} .
\end{equation}
Eq.~(\ref{eq:gamma}) is also valid without using the RWA, since the second term in the energy integral (\ref{eq:delta}) possesses no pole. The decay rate $\Gamma $ is also responsible for the additional broadening of the photoelectron peaks beyond the width given by  the pulse intensity envelope.

Equation~(\ref{eq:local_I}) can be solved explicitly
\begin{equation}
\label{eq:GSsol}
{a}_I(t)= e^{ \left(-i\Delta-\Gamma/2 \right)J(t)},
\end{equation}
where $J(t)=\int_{-\infty}^{t}g^2(t^\prime) dt^\prime $ grows smoothly with time. We find that the result (\ref{eq:GSsol}) is in excellent agreement with the exact calculations. It also simplifies greatly the computation of the spectrum. As it stands, ${a}_\varepsilon (t)$  in Eq.~(\ref{eq:CDE_F}) can be expressed as an integral of $a_I(t)$ \cite{Pahl99ZPD,Demekhin11SFatom}
\begin{equation}
\label{eq:solut}
{a}_\varepsilon (t)=-i \begin{array}{c} \left\{\frac{1}{2} d_\varepsilon \mathcal{E}_0 \right\} \end{array} e^{-i\delta t } \int_{-\infty}^t g(t^\prime)\, a_I(t^\prime)  \,  e^{i\delta t^\prime} dt^\prime,
\end{equation}
where we introduced the abbreviation $\delta=IP+\varepsilon-\omega=\varepsilon-\varepsilon_0$, which is the electron energy detuning from $\varepsilon_0$. Using now the explicit expression (\ref{eq:GSsol}) makes the computations of ${a}_\varepsilon (t)$ and of the spectrum $\sigma(\varepsilon)= \vert {a}_\varepsilon (+\infty) \vert^2 $  rather straightforward.

\begin{figure}
\includegraphics[scale=0.43]{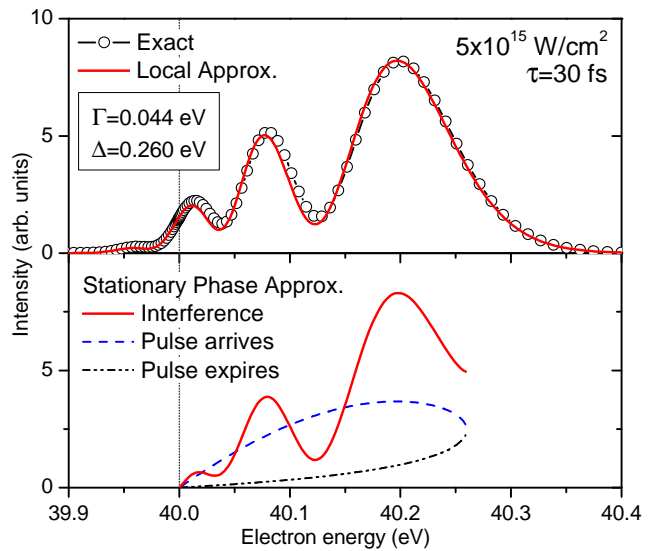}
\caption{(Color online) \emph{Upper panel:} The photoelectron spectrum of the hydrogen atom exposed to a Gaussian-shaped pulse of 30~fs duration, carrier frequency $\omega=53.6057$~eV and peak intensity $5\times 10^{15}$~W/cm$^2$ computed exactly (open circles) compared with the spectrum computed in the local approximation (solid curve) with the quantities $\Gamma=0.044$~eV and $\Delta=0.26$~eV. \emph{Lower panel:}  The spectrum obtained in the stationary phase approximation via Eq.~(\ref{eq:statphase}) with the same parameters (solid curve). The two individual contributions to the spectrum from the negative and positive times $t_s= \pm t_1(\varepsilon)=\pm \tau \sqrt{ \ln[\Delta/(\varepsilon-\varepsilon_0)]/2 }$  describing in Eq.~(\ref{eq:statphase}) the contributions when the pulse arrives and expires, respectively, are shown by broken curves. The energy dependent interference term in Eq.~(\ref{eq:statphase}) is proportional to $\cos\left\{-2(\varepsilon-\varepsilon_0) t_1 -\Delta [J(-t_1)-J(+t_1)]-\pi/2 \right\}$. In the energy interval $\varepsilon-\varepsilon_0 \in [0,\Delta]$  for which a stationary phase can be found, its argument accumulates a phase of $\Delta J(+\infty)$, which for a Gaussian pulse is equal to $\Delta \tau \sqrt{\pi/2}$. For the present parameters, the interference term accumulates a phase of $ 14.85\sim 2.36\cdot 2\pi $~rad, and, thus, the spectrum exhibits around two and a half oscillations of the intensity profile as seen in the figure. }\label{fig_approx}
\end{figure}

The photoelectron spectrum computed in the local approximation is depicted in the upper panel of Fig.~\ref{fig_approx}.  The AC Stark shift $\Delta=0.26$~eV and the decay rate $\Gamma=0.044$~eV computed via the above equations  (\ref{eq:delta}) and (\ref{eq:gamma}) have been utilized in the numerical calculations. It is seen that the spectrum calculated in the local approximation (solid curve) describes very well the exact spectrum (open circles).

Due to the success of the local approximation, we are able to gain deeper insight into the origin of the interference effects leading to the intensity modulation in the spectrum by evaluating analytically ${a}_\varepsilon (t)$. For this purpose we notice that the integrand in Eq.~(\ref{eq:solut}) contains a rapidly oscillating factor $exp \left[i\delta t^\prime-i\Delta J(t^\prime)\right]$ which is multiplied by a smoothly varying function. The main contributions to the integral stem from the times at which the phase $\Phi(t)=\delta_{\,}t-\Delta J(t)$ is stationary \cite{statphase}, i.e.  $\dot{\Phi}=0$. The resulting stationary phase condition,  $\Delta g^2(t_s) =\delta$, has a transparent physical meaning. It defines the time $t_s(\varepsilon)$ at which the energy of the ground state, continuously shifted by the AC Stark effect, moves across the energy position $\varepsilon$ of the dressed continuum state ($\delta=\varepsilon-\varepsilon_0$). For any pulse there are at least two stationary points for each value of $\varepsilon$: one, $t_1(\varepsilon)$, when the pulse is growing, and another, $t_2(\varepsilon)$, when it decreases. For a Gaussian pulse there are exactly two times,  $t_1(\varepsilon)=-t_2(\varepsilon)$.

\begin{figure}
\includegraphics[scale=0.4]{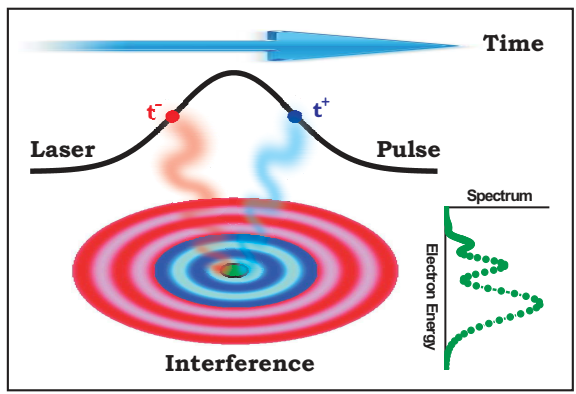}
\caption{(Color online) The intense laser pulse of sufficient high frequency to ionize the system by one-photon absorption induces a time-dependent AC Stark shift of the ground state of the system. Because of this dynamic shift, the photoelectron emitted along the pulse envelope has at every moment $t$  predominantly a specific kinetic energy $\varepsilon$ which is determined by that $t$ (at a given time $t$, the phase contributing to the integrand in Eq. (\ref{eq:solut}) is stationary at $\varepsilon= \varepsilon_0 + g^2(t) \Delta$, for details see text). Owing to the fact that the pulse envelope first grows and then falls, there are for a Gaussian pulse exactly two times (indicated as $t^-$ and $t^+$ in the figure) at which the emitted electron wave has the same energy $\varepsilon$. These two waves emitted with a time delay with respect to each other interfere. This dynamic interference gives rise to the strongly modulated intensity distribution of the photoelectron spectrum seen in Figs.~\ref{fig_spectra}-\ref{fig_approx}. The dynamic interference can be controlled by the intensity, duration, carrier frequency and shape of the pulse. A pulse envelope with more than one maximum leads to more than two waves which interfere and higher laser intensity to larger AC Stark shift $\Delta$ and to more oscillations of the spectrum. }\label{fig_scheme}
\end{figure}

By collecting in the integral (\ref{eq:solut}) these two main contributions at  $t_s=\pm t_1(\varepsilon)$, we obtain the following explicit approximate expression for the spectrum
\begin{equation}
\label{eq:statphase}
\sigma(\varepsilon)= \left| \frac{d_\varepsilon \mathcal{E}_0}{2}  \sum_{t_s=\pm t_1(\varepsilon)} g(t_s) e^{ -\Gamma/2 J(t_s)} e^{ i\left[ \Phi(t_s)\pm\frac{\pi}{4}\right]}    \right|^2 .
\end{equation}
The additional phase factors $\pm\frac{\pi}{4}$ result from higher terms in the expansion of the phase $\Phi(t)$ around the stationary points $\pm t_1(\varepsilon)$ computed for the Gaussian pulse.  The photoelectron spectrum evaluated via Eq.~(\ref{eq:statphase}) is depicted in the lower panel of Fig.~\ref{fig_approx} (solid curve). It is illuminating to see that an explicit simple expression reproduces nicely the exact spectrum (open circles in the upper panel of the figure) which is complicated to compute. The individual contributions to the spectrum of the two terms in the expression (\ref{eq:statphase}) are smooth and rather boring (broken curves in the lower panel of the figure).

Eq.~(\ref{eq:statphase}) allows one to uncover the physical origin of the strong modulations in the photoelectron spectrum. These are the results of the coherent superposition of two photoelectron waves emitted with the same kinetic energy at two different times. A cartoon visualizing this dynamic interference is shown in Fig.~\ref{fig_scheme}. This new physical effect causes enormous qualitative changes in the photoelectron spectrum which can easily be verified experimentally. Importantly, the interferences vanish if the AC Stark shift in the continuum vanishes (i.e., if $\Delta$ vanishes). Then, there are no stationary points along the pulse. We would like to stress that the dynamical interference can be controlled by the shape of the pulse envelope and is influenced by the choice of the carrier frequency. We are convinced that dynamic interference is an universal effect in ionization processes by high-frequency strong laser pulses and its experimental investigation is feasible. In the following we address these issues in some detail.

The predicted phenomenon also persists in other systems with more electrons. In the hydrogen atom discussed above the residual ion does not posses electrons and, therefore, the dynamic interference is governed by the AC Stark shift $g^2(t)\Delta $ which is a property of the ground state in the field. If the ion produced by photoionization does posses electrons, then this ion is obviously also subject to an AC Stark shift  $g^2(t)\Delta_{Ion}$. In this case, the dynamic interference pattern in the spectrum is governed by the total AC Stark shift $g^2(t)\Delta_{Tot}$  where $\Delta_{Tot}=\Delta-\Delta_{Ion}$  \cite{suplmat}. As the ion is usually harder to ionize than the neutral system, one can expect that $\Delta_{Tot}$ will be substantial even if both $\Delta$ and  $\Delta_{Ion}$ have the same sign. Moreover, depending on the photon energy $\omega$, the shifts  $\Delta$ and  $\Delta_{Ion}$  can have different signs, enhancing interference effects (see an explicit example in \cite{suplmat}). In many-electron systems, several electronic shells can be subject to the ionization, and different ionic states can be produced. Each ionic continuum will contribute to the shift of the ground electronic state leading all together to a final unique value of $\Delta$ at the carrier frequency $\omega$. In general, different ionic states $\vert F_\alpha \rangle$ will experience different AC Stark shifts $g^2(t)\Delta^\alpha_{Ion}$. The dynamic interference pattern in each of the partial electron spectra is determined by different total shifts $\Delta^\alpha_{Tot}=\Delta-\Delta^\alpha_{Ion}$  \cite{suplmat}, and one can expect dissimilar dynamic interference patterns in different partial photoelectron spectra. Experimental investigations of dynamic interference will be possible, in particular if the studied ionic thresholds are not too close to each other, i.e., if the resulting partial   spectra do not overlap. The AC Stark shift scales also with the probability to ionize the system. As photoionization cross sections usually decrease with the photon energy, the shift will be larger for frequencies $\omega$ which are not too far from the ionization threshold. In general, one can expect particularly large shifts, and hence more pronounced interference patterns, in systems with cross sections which vary strongly with the photon energy, e.g., in the vicinity of a shape resonance.

To demonstrate the feasibility of experiments on dynamic interference, we discuss the present status of FEL facilities. The currently operating FEL at FLASH \cite{FLASH} generates soft x-ray radiation in the energy range of  26--180 eV with flux of about $10^{13}$ photons per pulse with durations of 10--70 fs \cite{exp1}. Combined with appropriate focusing optics, peak irradiance levels of more than $10^{16}$~W/cm$^2$ can be achieved at present \cite{exp2}. The SASE FEL sources do not produce so far monochromatic radiation and their pulses consist of many spikes with randomly fluctuating properties. The impact of these problems on the experimental output has been studied, e.g., in Ref. \cite{exp3}. Nevertheless, the pulse properties at FLASH are continuously improving, and an average temporal coherence of 3--6 fs can already be achieved  \cite{exp4}. Moreover, several new techniques have been proposed for improvement \cite{exp5}, such as the selection of a single ultrashort radiation spike   \cite{exp6}. In addition, the recently tested FEL facility FERMI at Elettra \cite{exp7} operates in single-mode, providing coherent pulses (without spikes) with durations of 30 to 100 fs in the photon energy range of 12 to 413 eV, and it is expected to provide an unprecedented flux of about $10^{14}$ photons per pulse. Concluding, the pulse durations, photon energies, temporal coherence, and peak intensities necessary for experiments on dynamic interference are available at present.

In conclusion, intense high-frequency pulses give rise to a strong AC Stark effect in the continuum, which, in turn, modifies the ionization process dramatically. The effects are universal and can be verified experimentally by current techniques. Whenever matter is exposed to intense pulses, like in FEL experiments, ionization takes place, and the AC Stark effects in the continuum and the resulting dynamic interference will come into play. They will be essential for understanding the experimental results. The effects are thus both fundamental for light-matter interaction and of practical importance for the analysis of FEL experiments.

\begin{acknowledgments}
We thank A.I. Kuleff for many valuable discussions.
\end{acknowledgments}

\emph{Appendix}.--In this appendix we discuss how the full calculations were done. We have numerically solved exactly the system of equations which accounts for the interaction of all essential and nonessential electronic states of hydrogen with the laser pulse without employing the rotating wave approximation. For the peak intensities considered here, the effect essentially stems from the quantum motion of the electron eigenstates restricted to $ns$ and $nd$ Rydbergs with $n\le 6$, $mp$ Rydbergs with $m \le 16$, and, of course, $\varepsilon p$ continuum with $\varepsilon \le 100$~eV. Let us designate the time-dependent amplitudes for the population of the \textbf{even} $ns/nd$--states by ${a}_{e}(t)$, and for the \textbf{odd} $mp/\varepsilon p$--states by ${a}_{o}(t)$. The system of equations of motion of these states in the presence of the  pulse reads
\begin{equation} \nonumber
i\dot{a}_{e}(t)= E_{e} {a}_{e}(t) +\sum_{odd } \left(\frac{\mathcal{E}_0}{2} \left(d_{e}^{o} \right)^\dag  \right)g(t) \left(e^{i\omega t} +e^{-i\omega t} \right) {a}_{o}(t)
\end{equation}
\vspace{-0.5cm}
\begin{equation} \nonumber
i\dot{ {a}} _{o}(t)=  E_{o} {a}_{o}(t) + \sum_{even} \left(\frac{\mathcal{E}_0}{2} d_{e}^{o}  \right)  g(t)  \left(e^{i\omega t} +e^{-i\omega t} \right)  {a}_{e}(t)
\end{equation}
with boundary conditions $a_{1s}(-\infty)=1$ and $a_{i\ne 1s}(-\infty)=0$ for all other amplitudes. The continuous spectrum was discretized as $\vert F\varepsilon\rangle=\vert F\varepsilon_k\rangle \cdot \sqrt{\Delta\varepsilon_k} $, where $\vert F\varepsilon_k   \rangle$ represents the state in the continuum interval of $\Delta\varepsilon_k$ centered around the electron energy $\varepsilon_k$. The energy steps $\Delta\varepsilon_k$ were chosen linear with respect to $\sqrt{\varepsilon}$, and concentrated symmetrically around the central electron energy $\varepsilon _0 $. The convergence of the solution with respect to the electron eigenstates included, of the integration over time, and of the discrete representation of the continuous energy have been ensured.


\begin{thebibliography}{38}
\expandafter\ifx\csname natexlab\endcsname\relax\def\natexlab#1{#1}\fi
\expandafter\ifx\csname bibnamefont\endcsname\relax
  \def\bibnamefont#1{#1}\fi
\expandafter\ifx\csname bibfnamefont\endcsname\relax
  \def\bibfnamefont#1{#1}\fi
\expandafter\ifx\csname citenamefont\endcsname\relax
  \def\citenamefont#1{#1}\fi
\expandafter\ifx\csname url\endcsname\relax
  \def\url#1{\texttt{#1}}\fi
\expandafter\ifx\csname urlprefix\endcsname\relax\def\urlprefix{URL }\fi
\providecommand{\bibinfo}[2]{#2}
\providecommand{\eprint}[2][]{\url{#2}}


\bibitem{B1}
M.V. Fedorov,  \emph{Atomic and free electrons in a strong light field} (World Scientific, River Edge, 1997).

\bibitem{B2}
L. Allen and J.H. Eberly,  \emph{Optical resonance and two-level atoms} (Dover, Minneola, 1987).

\bibitem{B3}
C. Gerry and P. Knight,  \emph{Introductory quantum optics} (Cambridge U. P., Cambridge, 2004).

\bibitem{B4}
R. Loudon,  \emph{The quantum theory of light} (Oxford U. P., Oxford, 2000), 3rd ed.

\bibitem{A1}
N.B.\,Delone\,and\,V.P.\,Krainov,\,Phys.\,Usp.\,\textbf{42},\,669\,(1999).

\bibitem{A2}
W.  Happer, Prog. Quantum Electron. \textbf{1}, 51 (1970).

\bibitem{A3}
S.H.\,Autler and\,C.H.\,Townes,\,{Phys.\,Rev.} \textbf{100}, 703 (1955).

\bibitem{A4}
M. Haas,  \emph{et. al.}, { Am. J. Phys.} \textbf{74}, 77 (2006).

\bibitem{align}
H. Stapelfeldt and T. Seideman, {Rev. Mod. Phys}. \textbf{75}, 543 (2003).

\bibitem{shape}
B.J. Sussman,  \emph{et. al.},  {Science} \textbf{314}, 278 (2006).

\bibitem{birefringe1}
Ph.J.\,Bustard,\,\emph{et.\,al.},\,{Phys.\,Rev.\,Lett.}\,\textbf{104},\,193902\,(2010).

\bibitem{birefringe2}
A.V.  Sokolov and  S.E. Harris, {J. Opt. B: Quantum Semiclassical Opt.} \textbf{5}, R1 (2003).

\bibitem{control1}
S.A. Rice  and M. Zhao, \emph{   Optical Control of Molecular Dynamics }(Wiley, New York, 2000).

\bibitem{control2}
P.W.\,Brumer  and  M.\,Shapiro, \emph{ Principles of the Quantum Control of Molecular Processes} (Wiley, Hoboken, 2003).

\bibitem{FLASH}
W. Ackermann, \emph{et al.},  {Nature photonics} \textbf{1}, 336  (2007).

\bibitem{highharm1}
G. Sansone, \emph{et al.},  {Science} \textbf{314}, 443 (2006).

\bibitem{highharm2}
E. Goulielmakis,  \emph{et. al.},  {Science} \textbf{320}, 1614 (2008).

\bibitem{Sussman11}
B.J. Sussman,  { Am. J. Phys.} \textbf{79}, 477 (2011).

\bibitem{Pahl99ZPD}
E. Pahl,  \emph{et. al.}, {Z. Phys. D.} \textbf{38}, 215 (1999).

\bibitem{Chiang10}
Y.-C. Chiang,  \emph{et. al.},  {Phys. Rev. A} \textbf{81}, 032511 (2010).

\bibitem{Demekhin11SFatom}
Ph.V.  Demekhin and L.S. Cederbaum, {Phys. Rev. A } \textbf{83},  023422  (2011).

\bibitem{MolRaSfPRL}
L.S.\,Cederbaum,\,\emph{et.\,al.},\,{Phys.\,Rev.\,Lett.}\,\textbf{106},\,123001\,(2011).

\bibitem{DemekhinCO}
Ph.V.\,Demekhin,\,\emph{et.\,al.},\,{Phys.\,Rev.\,A.}\,\textbf{84},\,033417\,(2011).

\bibitem{DemekhinICD}
Ph.V.\,Demekhin,\,\emph{et.\,al.},\,Phys.\,Rev.\,Lett.\,\textbf{107},\,273002\,(2011).

\bibitem{Cederbaum81}
L.S.\,Cederbaum\,and\,W.\,Domcke,\,{J.\,Phys.\,B}\,\textbf{14},\,4665\,(1981).

\bibitem{Domcke91}
W. Domcke, {Phys. Rep.} \textbf{208},   97  (1991).

\bibitem{statphase}
N. Bleistein and R. Handelsman,   \emph{Asymptotic Expansions of Integrals.} (Dover, New York, 1975).

\bibitem{suplmat}
See Supplemental Material for this manuscript.

\bibitem{exp1}
K. Tiedtke, \emph{et al.}, New J. Phys. \textbf{11}, 023029 (2009).

\bibitem{exp2}
A.A. Sorokin, \emph{et al.}, Phys. Rev. Lett. \textbf{99}, 213002 (2007).

\bibitem{exp3}
N.\,Rohringer\,and\,R.\,Santra,\,Phys.\,Rev.\,A\,\textbf{77},\,053404\,(2008).

\bibitem{exp4}
S. Roling, \emph{et al.}, Phys. Rev. Special Topics - Accelerators and Beams \textbf{14}, 080701 (2011).

\bibitem{exp5}
J. Feldhaus, \emph{et al.}, J. Phys. B \textbf{38}, S799 (2005).

\bibitem{exp6}
E.L. Saldin,\emph{ et al.}, Opt. Commun. \textbf{212}, 377 (2002).

\bibitem{exp7}
http://www.elettra.trieste.it/FERMI/


\end{thebibliography}
\end{document}


\title{Supplemental Material\\ ~\\
Dynamic interference of photoelectrons produced by high-frequency laser pulses \\ ~}

\author{\firstname{Philipp~V.} \surname{Demekhin}}
\email{philipp.demekhin@pci.uni-heidelberg.de}
\affiliation{Theoretische Chemie, Physikalisch-Chemisches
Institut, Universit\"{a}t  Heidelberg, Im Neuenheimer Feld 229,
D-69120 Heidelberg, Germany}
\affiliation{Rostov State Transport University, Narodnogo Opolcheniya square 2,  Rostov-on-Don, 344038, Russia}

\author{\firstname{Lorenz~S.} \surname{Cederbaum}}
\email{lorenz.cederbaum@pci.uni-heidelberg.de}
\affiliation{Theoretische Chemie, Physikalisch-Chemisches Institut, Universit\"{a}t Heidelberg,
Im Neuenheimer Feld 229, D-69120 Heidelberg, Germany}

\pacs{33.20.Xx,  41.60.Cr, 82.50.Kx}

\maketitle

\section*{SEVERAL IONIZATION STEPS AND THRESHOLDS}

In the main manuscript we report on the theory of photoelectron spectra produced by laser pulses and their analysis. Numerically exact results were presented for the hydrogen atom which demonstrate the effect of dynamical interference beyond doubt. For transparency of presentation the equations were kept as needed for the hydrogen atom. Since this atom possesses only one electron, the produced ion cannot be further ionized. In this supplemental material we show how the present theory can be applied to systems with more electrons.

We start our discussion by first reminding the ansatz for the total wave function in the manuscript (Eq.~(1)) for the hydrogen atom
\begin{equation}
\label{eq:anzatz}
\Psi(t)= a_I(t)\vert I \rangle +\int  d\varepsilon \,a_\varepsilon (t) \vert  F\varepsilon \rangle e^{-i\omega t},
\end{equation}
where $a_I(t)$ and $a_\varepsilon (t)$  are the time-dependent amplitudes of the populations of the initial and final continuum states. When plugged into the Schr\"{o}dinger  equation, these populations could be expressed as (see Eqs.~(2b) and (3) in the manuscript and discussion therein)
\begin{subequations}
\label{eq:CDE}
\begin{equation}
\label{eq:CDE_I}
i\dot{a}_I(t)= \begin{array}{c}\left( \Delta-\frac{i}{2}\Gamma \right)\end{array}  g^2(t)\,{a}_I(t),
\end{equation}
\begin{equation}
\label{eq:CDE_F}
i\dot{ {a}}_\varepsilon (t)=  \begin{array}{c} \left\{\frac{1}{2} d_\varepsilon \mathcal{E}_0 \right\} \end{array}g(t) \,  a_I(t) + \left(IP+\varepsilon-\omega\right) {a}_\varepsilon (t).
\end{equation}
\end{subequations}
As explicitly shown in the manuscript, the quantum motion of the nonessential electronic states results in the time-dependent AC Stark energy shift, $\Delta g^2(t)$, of the ground electronic state (see Eq.~(\ref{eq:CDE_I})). Because of this dynamic shift, two photoelectron waves emitted at
rising and falling sides of the pulse superimpose, resulting in the dynamic interference effect. The interference pattern in the  photoelectron spectrum depends substantially on the shift $\Delta $: The larger the AC Stark shift the stronger is the effect and the more oscillations
appear in the spectrum (see the caption of Fig.~3 in the manuscript).

In the presence of a strong pulse, a system with more electrons can undergo further ionization. In order to incorporate this fact in the theory, one has to include the doubly-ionized electron continuum states in the ansatz (\ref{eq:anzatz}) for the total wave function. This is straightforwardly done. Following Refs. [S1-S6], the total wave function of the system as a function of time now reads
\begin{equation}
\label{eq:anzatzION}
\Psi(t)= a_I(t)\vert I \rangle +\int  d\varepsilon \,a_\varepsilon (t) \vert  F\varepsilon \rangle e^{-i\omega t}+\iint  d\varepsilon d\varepsilon^\prime \,b_{\varepsilon\varepsilon^\prime} (t) \vert G\varepsilon\varepsilon^\prime \rangle e^{-2i\omega t}.
\end{equation}
Here, $b_{\varepsilon\varepsilon^\prime} (t)$  is the time-dependent amplitude for the population of the doubly-ionized electron continuum state $\vert G\varepsilon\varepsilon^\prime \rangle$  which is already dressed by the field (i.e., by the energy of two absorbed photons $2\omega$), and   $\varepsilon^\prime$ is the kinetic energy of the secondary electron emitted by the ionization of the ion. Inserting $\Psi(t)$ into the time-dependent Schr\"{o}dinger  equation for the total Hamiltonian of the atom plus its interaction with the laser field, and implying the rotating wave [S7] and local [S8,S9] approximations, we obtain the following set of equations for the amplitudes (atomic units are used throughout)
\begin{subequations}
\label{eq:CDENEW}
\begin{equation}
\label{eq:CDENEW_I}
i\dot{a}_I(t)= \begin{array}{c}\left( \Delta-\frac{i}{2}\Gamma \right)\end{array}  g^2(t)\,{a}_I(t),
\end{equation}
\begin{equation}
\label{eq:CDENEW_F}
i\dot{ {a}}_\varepsilon (t)=  \begin{array}{c} \left\{\frac{1}{2} d_\varepsilon \mathcal{E}_0 \right\} \end{array}g(t) \,  a_I(t) + \left(IP+\begin{array}{c}\left[ \Delta_{Ion}-\frac{i}{2}\Gamma_{Ion} \right]\end{array}  g^2(t)+\varepsilon-\omega\right) {a}_\varepsilon (t),
\end{equation}
\begin{equation}
\label{eq:CDENEW_G}
i\dot{b}_{\varepsilon\varepsilon^\prime} (t)=  \begin{array}{c} \left\{\frac{1}{2} \widetilde{d}_{\varepsilon^\prime} \mathcal{E}_0 \right\} \end{array}g(t) \, {a}_\varepsilon (t) + \left(DIP+\varepsilon+\varepsilon^\prime-2\omega\right) b_{\varepsilon\varepsilon^\prime} (t).
\end{equation}
\end{subequations}
The last equation, Eq.~(\ref{eq:CDENEW_G}), describes the population of the doubly-ionized continuum states $\vert G\varepsilon\varepsilon^\prime \rangle$ by the subsequent ionization of the ion as a function of time. In this equation, the quantity $DIP=E_G-E_I$  is the double-ionization potential of the system, and $\widetilde{d}_{\varepsilon^\prime}$ is the   energy-dependent ionization dipole transition matrix element of the ion.

By comparing Eq.~(\ref{eq:CDENEW_F}) with Eq.~(\ref{eq:CDE_F}), one can see that the ionization potential is subject to a time-dependent AC Stark shift $\Delta_{Ion} g^2(t)$. More precisely, the ionic threshold is shifted by the AC Stark effect. If the energy of a single-photon is large enough to ionize the ion, this shift is induced by the quantum motion of the nonessential doubly-ionized continuum states. If the photon energy $\omega$ is insufficiently large to further ionize the ion, this shift also exists and is induced by the indirect coupling to the discrete and also continuum spectrum of the ionized system. In the case where the further ionization of the ion is possible, the energy of the ionic state is also augmented by the imaginary term $-\frac{i}{2}\Gamma_{Ion} g^2(t) $ seen in Eq.~(\ref{eq:CDENEW_F}). This term is responsible for the losses of the population of the ion by further ionization into all final doubly-ionized continuum states $\vert G\varepsilon\varepsilon^\prime \rangle$. The situation is similar to that discussed in Eq.~(\ref{eq:CDE_F}) (or equivalently Eq.~(3) in the main manuscript) where the ground state population decreases by ionisation losses. The expressions for $\Delta_{Ion} $  and  $\Gamma_{Ion} $  are analogous to those of the neutral system given in the manuscript, but for the ion.

Let us now discuss the consequences of the above and compare Eqs.~(\ref{eq:CDE_I},\ref{eq:CDE_F}) and Eqs.~(\ref{eq:CDENEW_I},\ref{eq:CDENEW_F}). Because of the AC Stark shift $\Delta_{Ion} g^2(t)$ of the ionic state, the dynamic interference pattern in the photoelectron spectrum is determined now by the total AC Stark shift $\Delta_{Tot}=\Delta-\Delta_{Ion}$. As the ion is usually harder to ionize than the neutral system, one can expect that $\Delta_{Tot}$ will be substantial even if both $\Delta$ and  $\Delta_{Ion}$ have the same sign. Moreover, depending on the photon energy, the shifts $\Delta$ and  $\Delta_{Ion}$ can have different signs. This may enhance the dynamic interference effect in the photoelectron spectrum. This situation
can indeed be achieved systematically by choosing carrier frequencies below the ionization potential of the ion. In this case, a nonresonant coupling of the ionization threshold with ionic Rydberg states may result in a negative AC Stark shift  $\Delta_{Ion}$ of the ion.

Until now we have discussed the situation of a single ionization threshold. A many-electron system generally exhibits several open ionization thresholds, and different ionic states can be produced by the photoionization. It is straightforward to incorporate this fact in the theory. One has just to sum over the final ionic states. The ansatz for the total wave function (\ref{eq:anzatzION}) now reads
\begin{equation}
\label{eq:anzatzNEW}
\Psi(t)= a_I(t)\vert I \rangle + \sum_\alpha \int  d\varepsilon_\alpha \,a_{\varepsilon_\alpha} (t) \vert  F_\alpha\varepsilon_\alpha \rangle e^{-i\omega t}+\sum_{\alpha\beta}\iint  d\varepsilon_\alpha d\varepsilon_\beta^\prime \,b_{\varepsilon_\alpha\varepsilon_\beta^\prime} (t) \vert G_\beta\varepsilon_\alpha\varepsilon_\beta^\prime \rangle e^{-2i\omega t},
\end{equation}
where the sums run over several final ionic continuum states  $\vert F_\alpha\varepsilon_\alpha \rangle$ and also doubly-ionized continuum states $\vert G_\beta\varepsilon_\alpha\varepsilon_\beta^\prime \rangle$.
The derivation of the equation of motion is in analogy to those for a single ionization threshold in Eqs.~(\ref{eq:CDENEW}) above. Obviously, each of the ionic continuum states $\vert F_\alpha\varepsilon_\alpha \rangle$  will contribute by a partial shift $\Delta^\alpha$ to the unique AC Stark energy shift $\Delta$ of the ground electronic state:
\begin{equation}
\label{eq:shiftNEW}
\Delta  =\sum_\alpha \Delta^\alpha  .
\end{equation}
In addition, each ionic state will experience an AC Stark shift  $\Delta_{Ion}^\alpha$, which is a sum over the partial shifts $ \Delta_{Ion}^{\alpha\beta }$ associated with a given doubly-ionized continuum state $\vert G_\beta\varepsilon_\alpha\varepsilon_\beta^\prime \rangle$:
\begin{equation}
\label{eq:shiftNEW_ion}
\Delta_{Ion}^\alpha =\sum_\beta \Delta_{Ion}^{\alpha\beta }.
\end{equation}

In general, different ionic states will experience different AC Stark shifts $\Delta_{Ion}^\alpha$. The dynamic interference pattern in each of the partial photoelectron spectra with photoelectrons of kinetic energy $\varepsilon_\alpha$ will be determined by a different total shift $\Delta^\alpha_{Tot}=\Delta-\Delta_{Ion}^\alpha$, and one can expect dissimilar interference pattern in different partial photoelectron spectra. Experimental investigations of systems with several open ionization thresholds would provide important information on the AC Stark shifts of different ionic states. The investigations will, of course, be possible if the ionic thresholds are not too close to each other such that the studied partial photoelectron spectra do not overlap.

In conclusion, in many-electron systems each partial photoelectron spectrum $\alpha$ will exhibit its own dynamic interference pattern dictated by $\Delta^\alpha_{Tot}=\Delta-\Delta_{Ion}^\alpha$ similar to the hydrogen atom case studied explicitly in the main manuscript.

\section*{DYNAMIC INTERFERENCE IN He}

\begin{figure}
\includegraphics[scale=0.5]{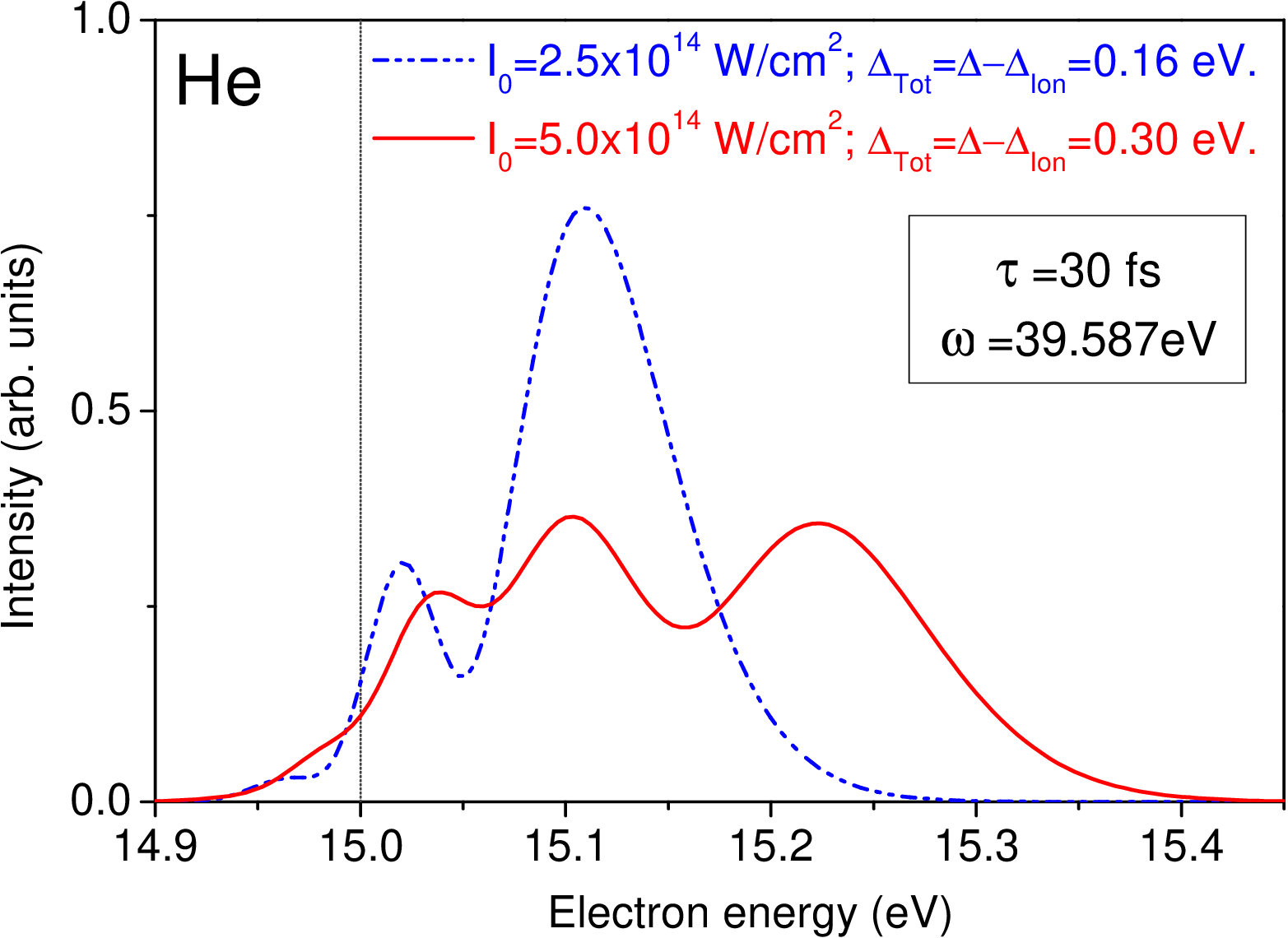}
\caption{(Color online) Photoelectron spectra of He exposed to a Gaussian-shaped pulse of 30 fs duration with carrier frequency $\omega = 39.587$~eV computed for two peak intensities (indicated in the legend). The photon energy $\omega $ is above the ionization potential of the neutral atom, but below the ionization potential   of the He$^+$ ion, and the ion cannot be ionized. This frequency is just below the first 2s/2p-Rydberg states of the ion (40.813 eV), leading to a negative AC Stark shift $\Delta_{Ion}<0$ for the ion He$^+(1s^1)$ state. The shift $\Delta$ of the neutral He is positive, and, hence, the resulting total shift in the photoelectron spectrum $\Delta_{Tot}=\Delta-\Delta_{Ion}$ (indicated in the legend for each intensity) is positive and enhanced.    }\label{fig_supplmat}
\end{figure}

We have exemplified our argumentation by as accurate as possible numerical calculations on the He atom. The calculation of the dipole transition matrix elements in the neutral and ionized He were performed in the Hartree-Fock approximation (which is, of course, exact for the He$^+$ ion). The photoelectron spectra of He computed at two different intensities  of the pulse with 30 fs duration are depicted in Fig.~\ref{fig_supplmat} of this supplemental material. The photon energy $\omega = 39.587$~eV  is above the ionization potential $IP=24.587 $~eV of the neutral He atom, but below the ionization potential $IP_{Ion}=54.418$~eV of the He$^+$ ion, and the ion cannot be ionized. This frequency is also just below the first 2s/2p-Rydberg states of the ion (40.813 eV). This results in a negative AC Stark shift $\Delta_{Ion}<0$ for   He$^+(1s^1)$ state (ionic threshold), which  stems mainly from the indirect coupling with the He$^+(2p^1)$ state via the field.  The shift $\Delta$ of the neutral He is positive and hence the resulting total shift in the photoelectron spectrum  $\Delta_{Tot}=\Delta-\Delta_{Ion}$  is larger than  that  for   neutral He. For the intensity of  $2.5\times10^{14}$~W/cm$^2$ the total shift amounts to $\Delta_{Tot}=0.02-(-0.14)=0.16$~eV and for the intensity of  $5\times10^{14}$~W/cm$^2$  to $\Delta_{Tot}=0.04-(-0.26)=0.30$~eV.   As a consequence, the computed electron spectra are shifted to   higher   energy   from the   photoelectron energy of $\varepsilon_0=\omega-IP=15$~eV (indicated by vertical dotted line) one would have at weak intensity. One can see from this figure  that the computed photoelectron spectra of the neutral He exhibit distinct multipeak patterns, which are due to the dynamic interference effect. We also notice that in He relatively low peak intensities of a few $10^{14}~\mathrm{W/ cm}^2$ are sufficient  to arrive at substantial effects which can be verified experimentally  by current FEL sources.

\makeatletter
\renewcommand*{\@biblabel}[1]{[S#1]}
\makeatother